\documentclass[aps,prb,twocolumn,showpacs]{revtex4}

\usepackage{graphicx}
\DeclareGraphicsExtensions{.png,.jpg,.eps}

\usepackage{xcolor}
\usepackage{color}
\usepackage{amsmath}
\usepackage{url}

\newcommand{\bluemark}[1] {#1}

\begin{document}

\title{Quantum Anomalous Hall effect in graphene coupled to  skyrmions}

\author{J. L. Lado$^1$, J. Fern\'andez-Rossier$^{1,2}$}

\affiliation{$^1$International Iberian Nanotechnology Laboratory (INL), Av. Mestre Jos\'e Veiga, 4715-330 Braga, Portugal }

\affiliation{$^2$Departamento de Fisica Aplicada, Universidad de Alicante, San Vicente del Raspeig, Alicante E-03690, Spain }


\date{\today} 

\begin{abstract} 
Skyrmions are topologically protected  spin textures, characterized by  a topological winding number $N$, 
 that occur spontaneously in some magnetic materials.
Recent experiments have demonstrated the capability to grow  graphene on top Fe/Ir, a system that exhibits
  a two dimensional Skyrmion lattice.  Here we show that  a weak exchange coupling between the Dirac electrons in graphene and a two dimensional Skyrmion lattice with $N=\pm 1$  drives graphene into a quantum anomalous Hall phase, 
 with a band-gap in bulk, a  Chern number ${\cal C}=2 N$ and  chiral edge states with perfect quantization of conductance $G=2N \frac{e^2}{h}$. 
  Our findings imply that  the topological properties of the Skyrmion lattice can be  imprinted in the Dirac electrons of graphene.
  
\end{abstract}
\maketitle

\section{INTRODUCTION}

Graphene is a zero gap semiconductor in the brink of becoming a topological insulator. 
It is no coincidence that several predictions of  topological phases in two dimensional  systems are based on 
some small modification of the model that actually describes graphene, namely,  a tight-binding Hamiltonian for electrons in a honeycomb lattice at half filling. 
In  a seminal paper\cite{Haldane88},  Haldane proposed that a honeycomb
crystal with a suitable magnetic flux decoration would become insulator
with a quantized Hall response, without Landau levels.  The ground-breaking
proposal of the quantum Spin Hall phase by  Kane and
Mele\cite{Kane-Mele1,Kane-Mele2} is also based on graphene with spin-orbit
coupling, that is  mathematically related to the Haldane model.  A
QSH-like  phase was also predicted for graphene under the influence of
a perpendicular magnetic field and ferromagnetic order\cite{Abanin06}.    
In addition, the quantum
Hall effect has been observed  in graphene at a relatively  low magnetic
field at low temperatures\cite{Young14}, and even at room
temperature\cite{roomT} at high fields. 
Furthermore, the combination of exchange interaction and spin-orbit
coupling has been predicted to give rise to 
the QAH phase\cite{Qiao10}. 
Here we propose a topological phase that, unlike all of the above, requires no spin-orbit coupling and no applied magnetic field.

Topological insulating phases in two dimensions attract interest because of
the very special transport properties, such as the perfect conductance
quantization of  the quantum Hall phase\cite{VK}. These special transport
properties arise from the fact that  
 topological phases have a gapped bulk  and conducting edge states.  
Interestingly, QAH phases  have chiral edge states for which intraedge 
backscattering is impossible. Thus,  quantization of conductance of the 
QAH phase should be as perfect as the one observed in 
the quantum Hall effect.  This 
 contrasts with  QSH edge states, for which only  the total absence of 
time-reversal symmetry breaking impurities, such as nuclear spins or local 
moments, prevents  backscattering. As a result, the quantization of 
conductance in QSH  and QSH-like phases is far\cite{Konig07,Young14} 
from the perfection observed in graphene QH systems\cite{roomT,VK}.

  \begin{figure}[t!]
 \centering
 \includegraphics[width=0.45\textwidth]{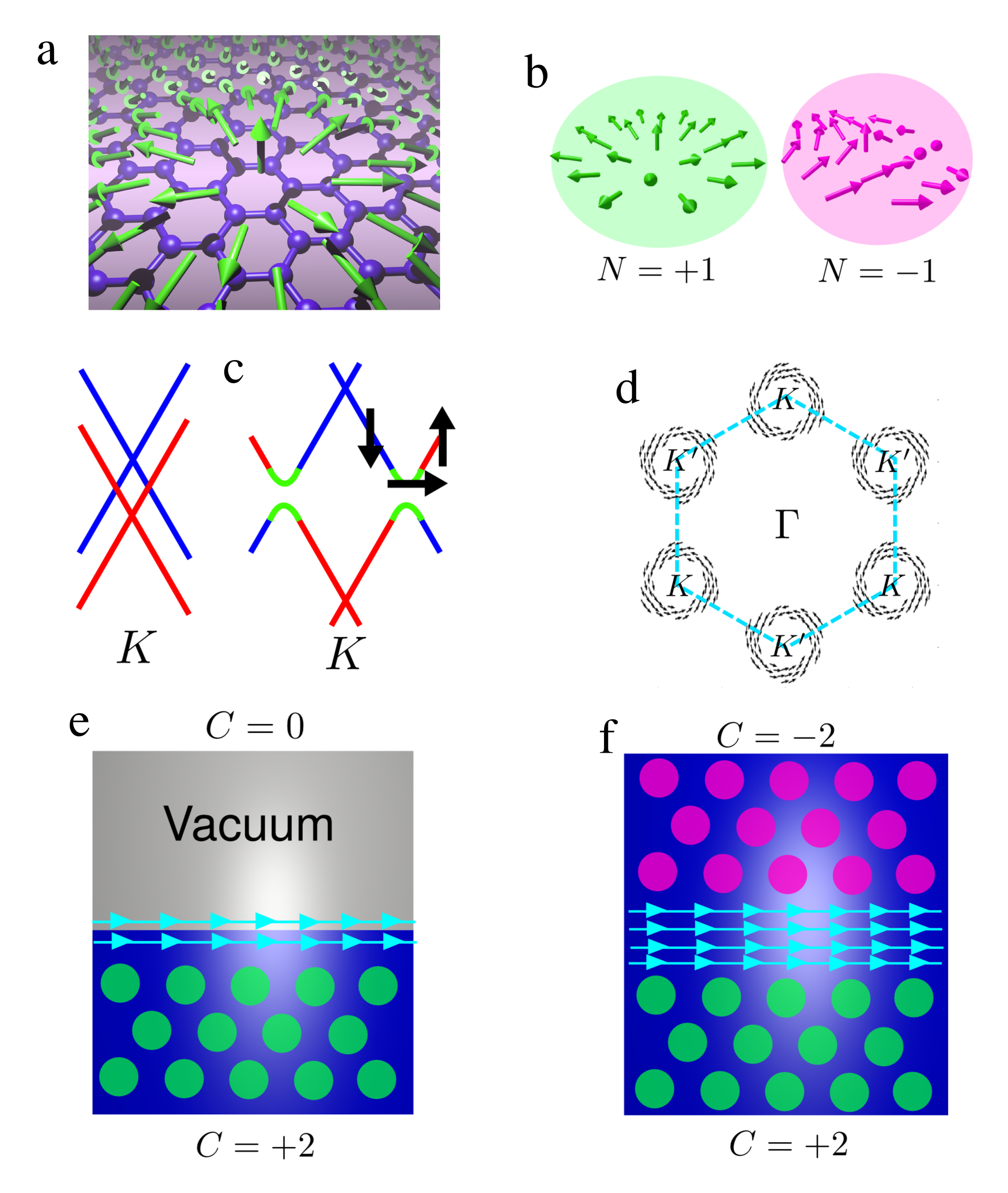}
\caption{ (a) Scheme of a graphene layer deposited on a skyrmion lattice (only the skyrmion cores are shown). (b) Skyrmions with
$N=\pm1$, that trigger  a QAH.
(c) Influence of the exchange to the skyrmions on the Dirac cones: spin splitting and band-gap opening
(d) Spin texture induced in the reciprocal space by the Skyrmion driven
gap opening. Edge states between vacuum and 2d Skyrmion crystal (e), and
between two skyrmionic crystals with different chirality (f). }
\label{fig1}
\end{figure}

The notion that exchange interaction with  the  local moments with 
non-collinear  spin textures affects severely the kinetic energy 
of itinerant electrons, goes back to the proposal of  the double-exchange 
mechanism\cite{Hasegawa}.  Later, the role 
of non trivial effects on the Berry phase 
was recognized\cite{Loss90,BerryDE}, and the notion 
that  non-collinear spin textures  induce an effective orbital 
magnetic field that
would  lead to an anomalous Hall term was put forward\cite{Millis99,Calderon01,Batista1,Batista2,Batista3,Taguchi2001,RMP2010}.  
It was also shown by  Ohgushi {\em et al.}\cite{Ohgushi2000} that 
the double exchange model on a two dimensional  Kagome lattice with 
non-collinear classical spins led to a quantum anomalous Hall
phase with quantized Hall conductance. 
Recent experiments have demonstrated the possibility of growing
graphene islands on top of a  two dimensional skyrmion lattice
hosted by a single atomic layer of   Fe deposited on an Ir(111)
substrate\cite{RW2014}. This experiment,  together with  the notion
that non-collinear spin textures induce very interesting effects on
conduction electrons, \cite{Loss90,BerryDE,Millis99,Taguchi2001,RMP2010,Ohgushi2000}, motivate
our study of graphene electrons coupled to a skyrmion lattice.  
Interestingly, we find that this system can exhibit the Quantum
Anomalous Hall effect (Fig. 1e,1f), a topological phase of matter that is
being actively pursued in condensed matter physics\cite{QAHE}

\bluemark{The paper is organized as follows. In Section II we present
the tight binding model to study the skyrmion proximity effect in graphene.
In section III  we show the non-trivial phases that arise
by proximizing graphene to a triangular and rectangular skyrmion
lattices respectively, in the weak coupling limit at half filling.
Finally, in section
IV we summarize our conclusions.}

\section{Model}

Skyrmions are non-collinear spin 
textures (Fig. 1a and 1b) characterized by a topological number\cite{Nagaosa-Tokura2013}
\begin{equation}
N= \frac{1}{4\pi}\int\vec{n}\cdot\left(\frac{\partial \vec{n}}{\partial x}
\times\frac{\partial \vec{n}}{\partial y}\right) d A
\end{equation}
where $\vec{n}=\frac{\vec{M}}{|\vec{M}|}$ is the unit 
vector associated to the skyrmion magnetization. 
Skyrmions are being very actively studied in 
the context of  spintronics\cite{Nagaosa-Tokura2013,Sampaio13}, and they 
have been found both in bulk compounds\cite{Nagaosa-Tokura2013}  and 
two dimensional systems \cite{Heinze2011}.
Here we propose a mechanism to induce the QAH phase in 
graphene, fully independent of the spin-orbit coupling of carbon.
Namely, it is based on exchange interaction with a magnetically ordered surface 
that hosts a skyrmion lattice (Fig. 1a).

Our starting point is the  tight-binding model
for electrons in graphene ${\cal H}_0 = \sum_{ij\sigma} c_{j\sigma}^\dagger c_{i\sigma}$
plus their exchange to an
arbitrary magnetization field, that is treated classically: 
\begin{equation}
{\cal H}= {\cal H}_0 + J \sum_{i} \vec{S}_i\cdot\vec{M}_i
\end{equation}
Here $i$ labels the sites of the honeycomb lattice,
 $J$ is the short-range exchange interaction constant, 
$\vec{S}_i$  and $\vec{M}_i=M_0 \vec n_i$ are 
the electronic spin density operator  and classical
magnetization, respectively, 
evaluated at  site $i$.   The length of the magnetization
field $M_0$ is assumed to be the same for all sites.   Unlike most of 
previous work\cite{nagaosa2015}, we focus on  the weak  coupling limit $J<<t$, adequate  for proximity induced magnetism. 
Therefore, the exchange
field act as a perturbation on the Dirac spectrum. 

The magnetization of  a single skyrmion  can be
expressed as a map from the plane described in
polar coordinates $r,\phi$ to the unit sphere, described in
spherical coordinates $\Theta, \Phi$, 
\begin{equation}
\vec{n}=  \left(\sin\Theta(r) \cos\Phi(\phi), \sin \Theta(r) \sin\Phi(\phi), \cos \Theta(r) \right)
\label{skymag}
\end{equation}
where $\Theta(r)$ is such that $\Theta(r=0)=0$ 
and $\Theta (r>R_{Sky})=\pi$ , where $R_{Sky}$ is the skyrmion
radius. $\Phi(\phi)= N \phi + \gamma$, where $N$ is the
skyrmion number that accounts for its
vorticity\cite{Nagaosa-Tokura2013} and $\gamma$ is a phase
that determines the helicity of the skyrmion. Since $\gamma$ can
be gauged by a rotation along the z-axis, we 
take $\gamma=0$ without loss of generality. 
\bluemark{For the sake of simplicity, in the following
we will assume a stepwise profile for the magnetization
\begin{equation}
\Theta^{\text{hard}}(r) =
\begin{cases}
0 & r=0 \\
\pi/2 & 0<r<R_{Sky} \\
\pi & r>R_{Sky} \\
\end{cases}
\end{equation}
We have verified that the results do not change qualitatively
if we use a  smooth parametrization of the azimuthal angle. 
}

  \begin{figure}[t!]
 \centering
 \includegraphics[width=0.5\textwidth]{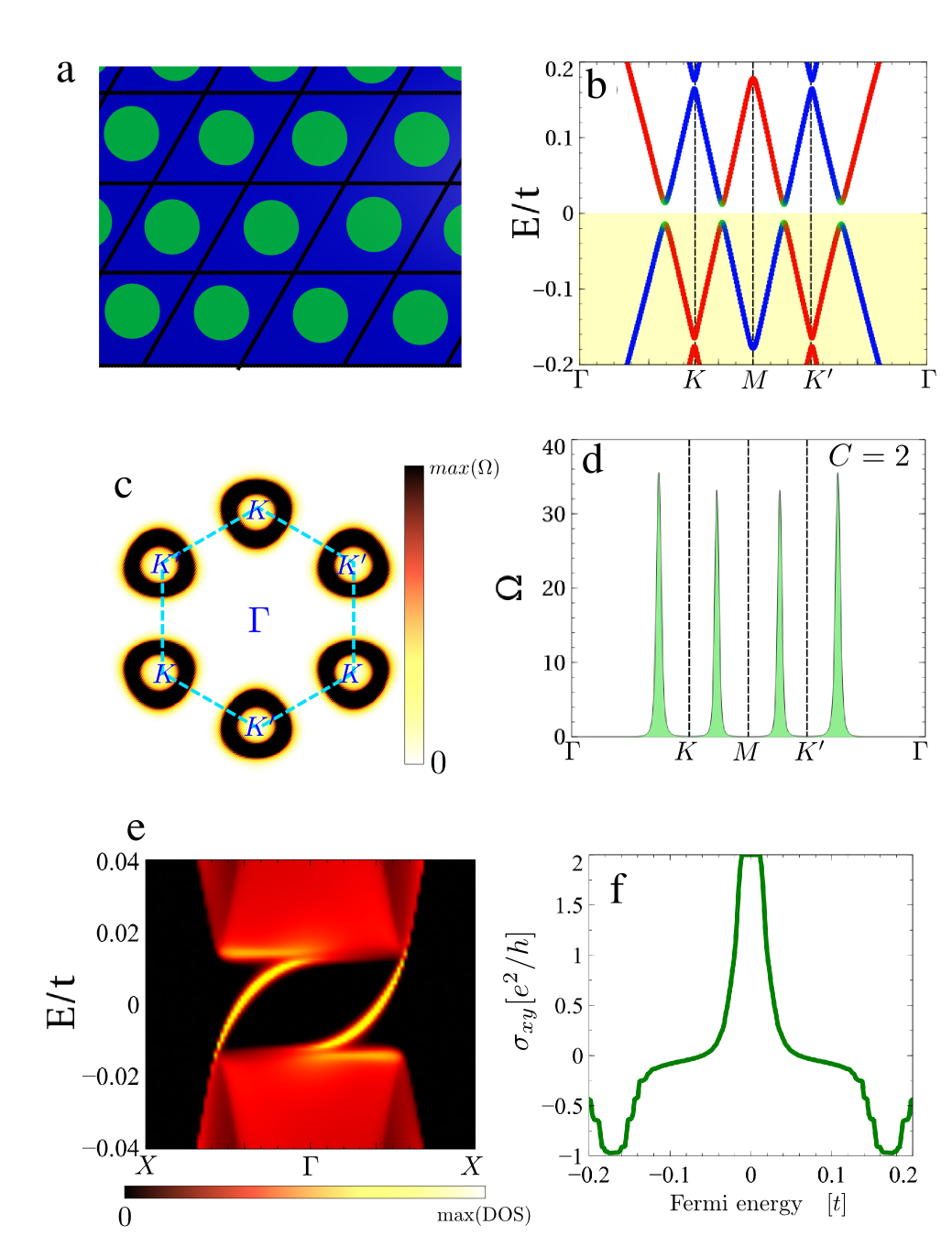}
\caption{(a) Scheme of a triangular arrangement
of skyrmions and (b) band structure of a 5x5 honeycomb supercell with a
skyrmion with winding number equal to 1 and $J = 0.3t$.
(c) Contour plot of the Berry
curvature, with a small trigonal
warping
around  the corners of the hexagonal Brillouin zone. 
(d) Positive Berry curvature  localized at the  region of
band-gap  opening.
(e) Surface density
of states on the
termination of a  semi-infinite graphene
plane, showing two in-gap 
chiral states. 
(f) Anomalous Hall conductivity as a function of Fermi energy
}
\label{fig2}
\end{figure}

\section{Quantum anomalous phase at weak coupling}

\bluemark{In this section we present the
result for the weak exchange limit $(J<<t)$. This
situation is the one to be realistically obtained for graphene
on top of a skyrmion lattice. We show that a topological gap
opens for arbitrarily small exchange coupling and independently on
the type of skyrmion lattice, triangular or rectangular.}

\subsection{Triangular skyrmion lattice}

We now consider the properties of graphene interacting with a triangular
crystal of skyrmions.
In Fig 2b we show the energy bands for a 
 5x5 supercell with
one skyrmion.
It is apparent that exchange interaction with the skyrmions
opens up a gap in graphene.  This can be understood as follows.
We write the exchange part of the Hamiltonian as
\begin{equation}
{\cal V}= + J \sum_{I} \vec{S}_I\cdot\left(\langle\vec{M}\rangle+ \delta\vec{M}_I\right)
\label{vint}
\end{equation}
where  $\langle \vec{M}\rangle = \frac{1}{N}\sum_{I} \vec{M}_I$ is the 
  average magnetization, and $\delta\vec{M}_I = \vec{M}_i - \langle \vec{M}\rangle$ are the fluctuations. Ignoring the fluctuation term, 
the average magnetization induces a spin-splitting, shown in figure
1b,  where the bands have a well defined spin along the average
magnetization.  The resulting  conduction band of  one spin projection
is degenerate with the valence band of the opposite spin projection, 
defining a circle of degenerate points that, at half
filling, happens to be the Fermi surface (see Fig. 1c).   
 For non-collinear spin textures,  $\delta \vec{M}$ has terms  orthogonal to 
 $\langle\vec{M}\rangle$,  $\delta \vec{M}_{\perp}$, that  open up the gap at the
degeneracy circle, \bluemark{and creates a spin vorticity in the
reciprocal space (see Fig. 1d)}.  A similar argument has been used by
Qiao and coworkers in their proposal\cite{Qiao10} for QAH in graphene with Rashba
spin-orbit coupling and a exchange field. 
In our system the $\delta\vec{M}_{\perp}$ part of the Hamiltonian plays the same
  role as the
Rashba coupling in their model. However, we found that only skyrmions with  topological charge $N=\pm1$ are able to open
a topological gap.

  \begin{figure}[t!]
 \centering
 \includegraphics[width=0.5\textwidth]{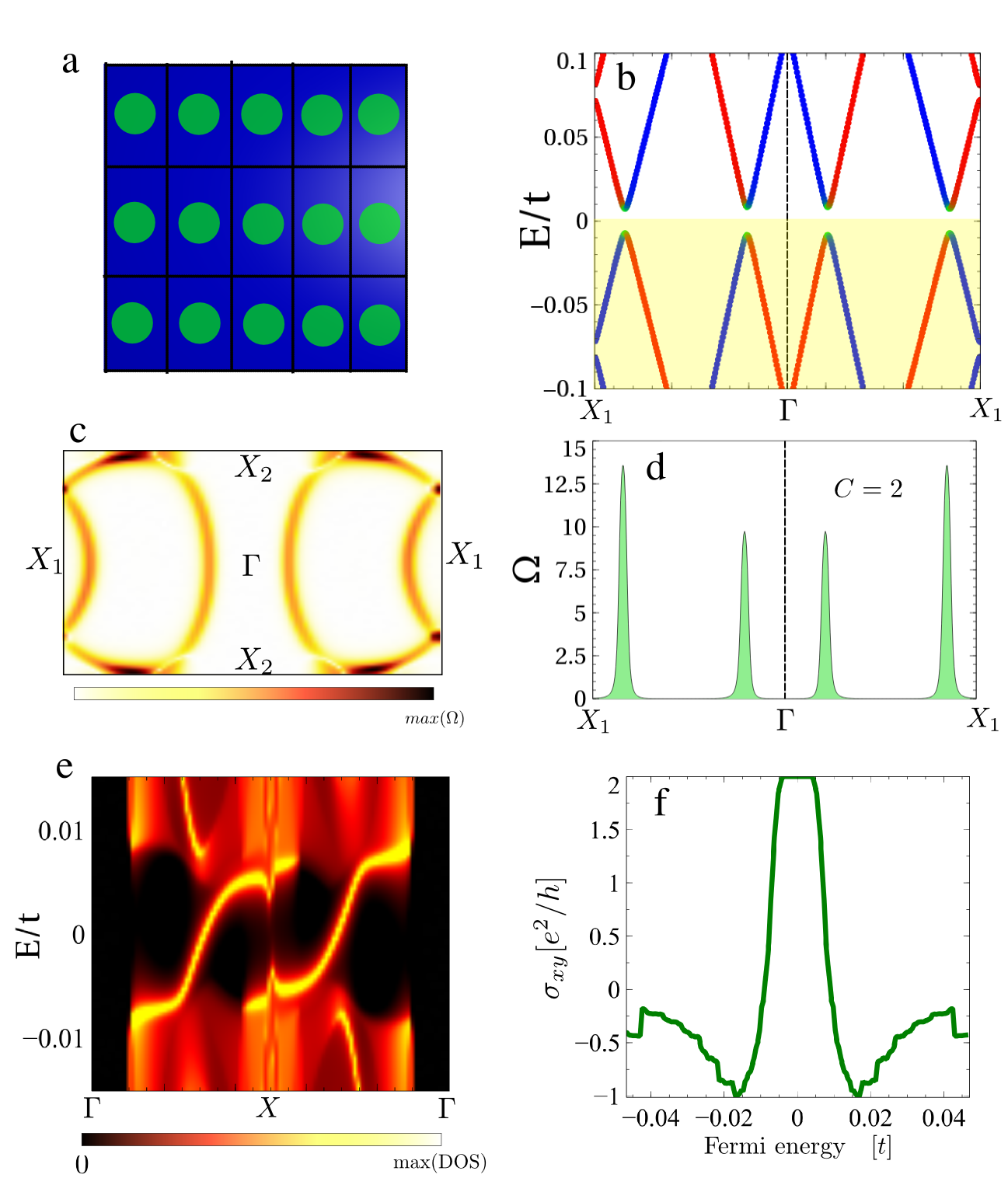}
\caption{  (a) Scheme of a rectangular skyrmion
crystal. (b) Band structure of half filled graphene
with a rectangular unit cell, coupled to 
the rectangular skyrmion lattice with $J=0.3t$. (c) Berry curvature in the whole
Brillouin zone, showing a same sign behavior which sums up to
a ${\cal C}=+2$ total Chern number. (d) Berry curvature along the $k$ path shown in
the band structure, showing a non-vanishing contribution in the
anti-crossings points. (e) Surface DOS in a seminfinite geometry showing
two chiral states. 
(f) Anomalous Hall conductivity as a function of Fermi energy
}
\label{fig3}
\end{figure}

  The non-trivial nature of the  gap  of Fig. 2b can be anticipated from 
the Berry curvature\cite{RMP-Xiao} profile  shown in Fig 2d.   A non vanishing Berry curvature arises at the crossing circle
of the spin-split Dirac points \bluemark{(see Fig. 2c)}, where the spin-flip terms open-up a gap.     

The Hall conductivity can be  be expressed as an integral of the Berry curvature\cite{Haldane88,RMP2010,TKNN}:
\begin{equation}
\sigma_{xy} (E_F) =\frac{1}{2\pi}\frac{e^2}{h}\sum_n \int f(\epsilon_n(\vec{k}),E_F) \Omega_n(\vec{k})d^2k 
\label{Chern}
\end{equation}
where $f(\epsilon_n(\vec{k}))$ is the Fermi occupation function, and $n$ labels the bands. When the Fermi energy, $E_F$, lies inside a gap, the Hall conductivity is proportional to a Chern number ${\cal C}$\cite{TKNN}, which is an integer number ${\cal C}$, resulting in a  quantized Hall conductivity 
$\sigma_{xy}=\mathcal{C} e^2/h$. 
Our calculations show that when $E_F$ lies in the gap opened by the coupling to the $N=\pm 1$ skyrmions,
the  Chern number is given by. 
\begin{equation}
{\cal C}=2N
\end{equation}
This is the central result of this work:  the topological winding number of the skyrmions is imprinted into  the Dirac electrons.  Several consequences follow. 
First,    two  chiral edge states are expected to occur at the boundaries of the crystal.  This is confirmed by our calculations, using a recursive Green
function method\cite{edges} to calculate the surface states, as shown in Fig. 2e.   Second,   an interface
between two  skyrmion  lattices, with opposite
skyrmion numbers $N=1$ and $N=-1$,  is expected to
show 4 interface states according to the index
theorem (see fig. 1f).  Third, when $E_F$  lies inside the gap, the Hall conductance
is quantized \cite{TKNN,RMP2010} $\sigma_{xy}= {\cal C}\frac{e^2}{h}$. 
Importantly,  the  Hall conductivity also takes large, but not-quantized values,  when  effect  Fermi energy lies close to, but outside, the 
gap (see Figs. 2f, 3f),  on account of the finite Berry curvature
integrated over the occupied states\cite{Haldane04}. 

\subsection{Square skyrmion lattice}

We now address the question of the influence of the type of skyrmion lattice on the existence
of the QAH phase. In particular,  motivated by
experimental results\cite{RW2014}, 
we consider a rectangular skyrmionic crystal, commensurate with
the graphene lattice (see figure 3a). 
We find that the topological character of such system strongly depends on whether
the valley mixing is an important effect.

The topological
phase is observed for those unit cells in which the two valleys fold
to different points in the Brillouin zone, which avoid
intervalley mixing. 
\bluemark{As in the triangular case, the in-plane components open up
a gap in the exchange split bands (Fig. 3b), which leads
to a finite non-vanishing Berry curvature localized in the band crossing points (Fig. 3d).} 
\bluemark{The Fermi surface in this case can be more complex than in the triangular skyrmion lattice due to the
large folding of the reciprocal
space, which is also reflected
in Berry curvature {Fig. 3c}. The calculated Chern number is also $C=2N$,
as in the triangular lattice, which in a semi-infinite geometry
gives rise to two copropagating edge branches (Fig. 3e).}

\bluemark{Away from half filling, when the Fermi energy no longer
lies within the gap, the anomalous response is still non-vanishing up to
energies 4 times the gap (Fig. 3f). The sign of the anomalous response is again the
same for electrons and holes as in the triangular case. In comparison, upon
entering the conduction (valence) band, the response can rapidly
become negative and with a value close to 1, due to the
presence of an edge branch located in the
valence (conduction) band (see Fig. 3e), that coexist with normal
valence (conduction) states.}

\subsection{Experimental verification}

The experimental verification of
our proposal is not far from the  state of the art.   
The optimal conditions to detect perfect edge transport associated to the  QAH phase
is induced by a skyrmion lattice hosted by an insulating magnetic
material that couples to the graphene electrons.  Three
ingredients  have been demonstrated in different systems.  First,  recent
experimental results showing the possibility of growing graphene on the
surface of Fe/Ir, a surface that hosts a skyrmion
lattice\cite{RW2014}.   Second, and independent from the first, recent
experiments show that non-quantized anomalous Hall  is induced in graphene
by proximity to a ferromagnetic insulating
substrate \cite{Wang15}.    Finally,  skyrmion lattices have been observed
in insulating  chiral-lattice magnet Cu$_2$OSeO$_3$\cite{Seki12}.  

The critical figure of merit to realize our proposal is the magnitude of the skyrmion induced gap.
This  is controlled by the
strength of the exchange field of the underlying magnetic state.
To gain some insight of the gap opening mechanism,
it is convenient to change independently the off-plane
(exchange shift) and in-plane (spin mixing) components of the
skyrmion texture. 
Our numerical calculations 
(see Fig. 4b)  for the triangular  lattice  of  hard core skyrmions ($n_z(r<R_{Sky}) = 0$) 
 show that the gap  satisfies 
$\Delta \propto J_{in} J_{off}/t$
 where $J_{in}$ and $J_{off}$ are the magnitude of the in-plane
and off-plane components of the exchange field (see Fig. 4b,4c,4d). In the case of
constant exchange strength, a quadratic exchange dependence is obtained
$\Delta \propto J^2/t$.  

For reference, and given that there are no measurements of J,  we take
it from DFT calculations for graphene on top of
BiFeO$_3$ \cite{qah-af} that give $J=70$ meV, and $t=2.6$ eV.
With the previous parameters, the
topological gap would get a value of $\Delta \simeq 0.4$ meV, within reach of
transport spectroscopy.
Taking these numbers, the  window  of $E_F$  within which the non-quantized intrinsic Hall effect
would be sizable extends
in a window of $1-3$ meV around the Dirac energy. 
Our calculations indicate that the non-quantized Hall conductivity would be
larger in the case of the square lattice (Fig. 3f).  We also note that the sign of the 
skyrmion-induced Hall contribution would be the same for electrons
and holes, in contrast with the conventional Hall effect.


  \begin{figure}[t!]
 \centering
 \includegraphics[width=0.5\textwidth]{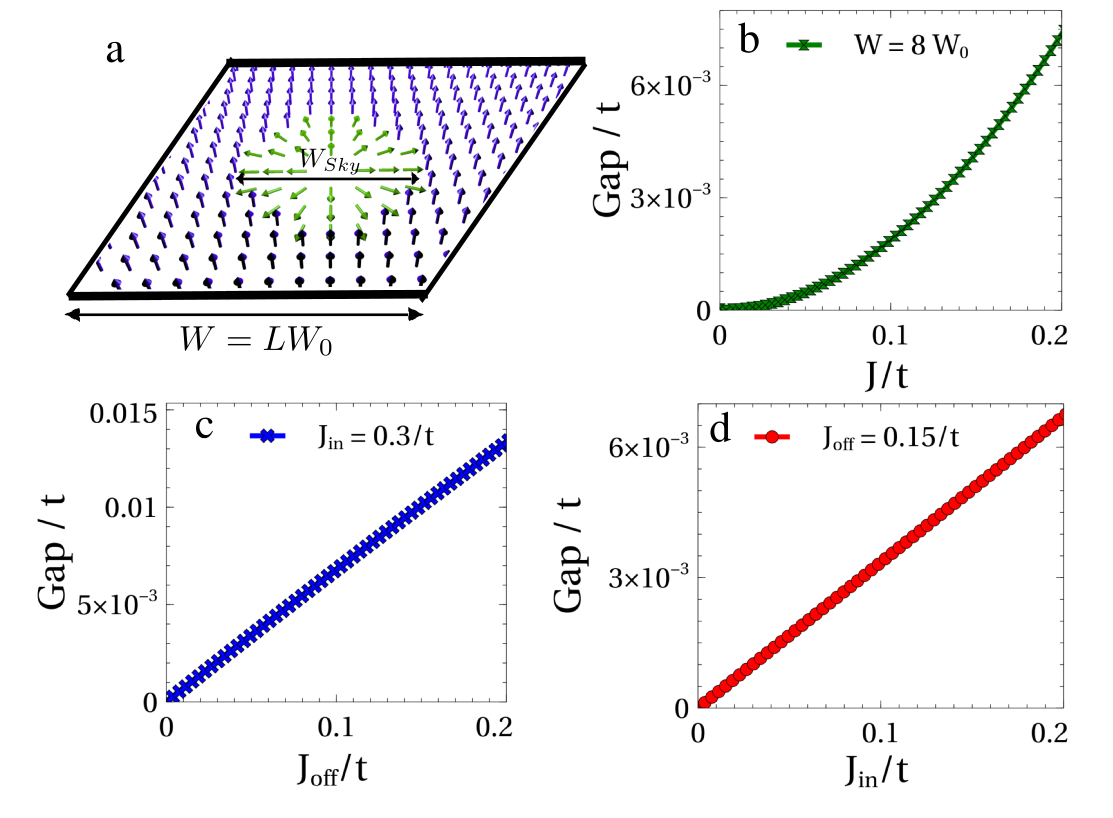}
\caption{ (a) Scheme of a hard core skyrmion, showing an in-plane
radial magnetic profile and an out-core off-plane magnetism.
Evolution of the topological gap at half filling with the global exchange (b),
only the out core exchange (c) and the core exchange (d). 
}
\label{fig4}
\end{figure}

\section{CONCLUSION}


We have shown that, at half filling,  graphene  with a weak exchange coupling  to a skyrmion lattice   develops a quantum anomalous
Hall phase, with gapped bulk and chiral edge states that should have perfect quantization. 
This  occurs at least for two different skyrmion lattices, rectangular and triangular,  and seems a generic feature
as long as the skyrmion lattice does not produce valley mixing.
The Chern invariant ${\cal C}$ that characterizes the QAH phase is given by the topological invariant that describes the individual skyrmions $N$,  through the remarkable relation ${\cal C}= 2N$, valid for $N=\pm 1$.  Thus, graphene  edges will have 2 chiral edge states, and  graphene on top of an interface between two  skyrmions lattices  with opposite skyrmion number $N=+1$ and $N=-1$ will have 4 chiral edge states. 
Our proposal is different from previous proposals of QAH phase in graphene because it requires no spin-orbit coupling and no magnetic field acting on the graphene electrons. 

{\em Note added:} During the final  stages of the completion  of this manuscript, we became aware
of a work predicting a Quantum Anomalous phase for electrons
on a square lattice coupled to  skyrmions \cite{nagaosa2015} in the strong coupling limit ($J>>t$).


\section{Acknowledgments}

JFR acknowledges  financial support by MEC-Spain  (FIS2013-47328-C2-2-P)   and Generalitat Valenciana (ACOMP/2010/070), Prometeo. 
This work has
been financially supported in part by FEDER funds.  We acknowledge
financial support by Marie-Curie-ITN 607904-SPINOGRAPH. J. L. Lado thanks
the hospitality of the Departamento de Fisica Aplicada
at the Universidad de Alicante. We thank L. Brey and
D. Jacob for useful discussions.

%

\appendix

\section{Strong coupling limit}

  \begin{figure}[t!]
 \centering
 \includegraphics[width=0.5\textwidth]{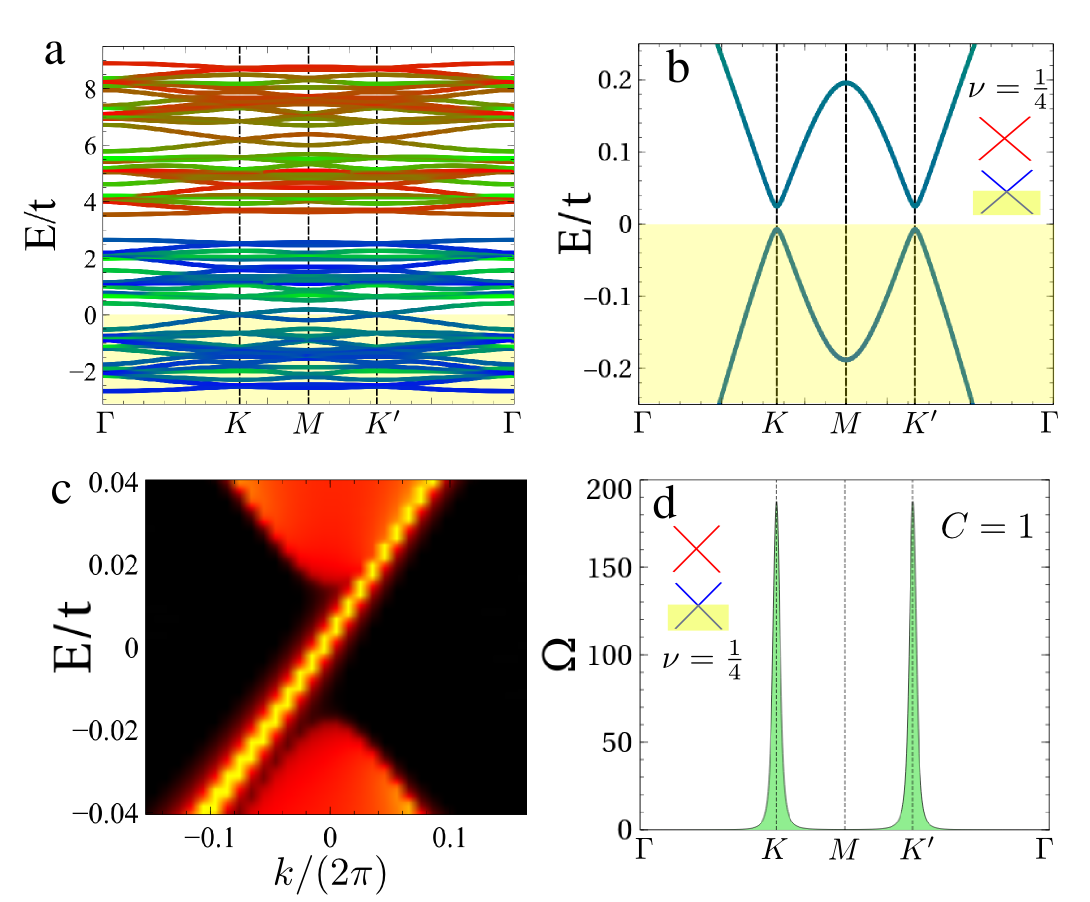}
\caption{Graphene supercell over a triangular
skyrmion crystal.
Band structure
(a,b) at filling $1/4$ in the large exchange
limit (J=3t). As shown by the
surface density of states (c) and the Berry curvature (d), the system develops a quantum anomalous
Hall state with a total Chern number $\mathcal{C}=1$ At large
exchange, the half filling state becomes trivial, whereas
the filling 1/4 becomes topological resembling the Haldane model.
}
\label{large}
\end{figure}

We   discuss here  the behavior of the system in the large $J$ limit, where $J>>t$
and the spin-splitting of the bands is much larger than the bandwidth. This strong coupling
limit has been  considered in previous works\cite{Ohgushi2000}, but is not realistic in the case of graphene.
In the strong coupling limit at half filling, the system behaves as a topologically trivial magnetic insulator, where both valence and conduction states of the majority spin
are completely full.  Since spin degeneracy is completely lifted in this case,  the Fermi energy  lies at the Dirac point  at quarter filling.  In this case   (Figs. \ref{large}), the low energy states are described by a
gaped  Dirac like spectrum and  our calculations in  this limit   give ${\cal C}=1$, for $N=1$. Therefore, 
this is   topologically different to the weak-coupling case discussed in the main text,  and much closer to the  original QAH phase proposed by  Haldane \cite{Haldane88}. 

The physical origin of this topological phase can be understood as follows. 
 The standard\cite{Millis99,RMP2010,Taguchi2001,Ohgushi2000} spin rotation is performed, so that the exchange term is always diagonal\cite{BerryDE,Millis99}. The local nature of this transformation generates a coupling to a  gauge field in the hopping operator\cite{Millis99,RMP2010,Taguchi2001,Ohgushi2000} that describes an effective inhomogeneous magnetic field, that is responsible of the band-gap opening in this limit. 
 In
that limit, a skyrmion with topological number
$N$ creates an effective field equal to
$N\Phi_0$, where $\Phi_0=e/h$ is the magnetic
flux quantum  \cite{Nagaosa-Tokura2013}.

  \begin{figure}[t!]
 \centering
 \includegraphics[width=0.5\textwidth]{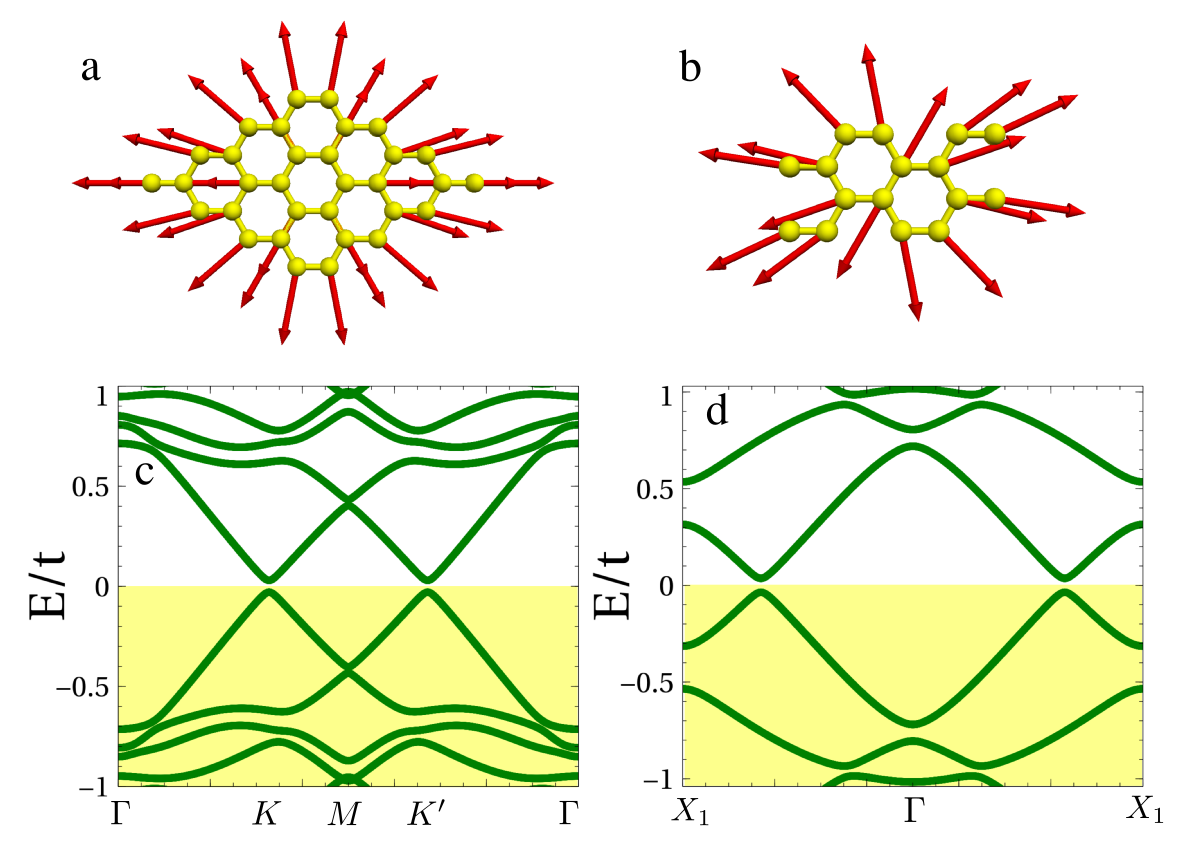}
\caption{Unit cells (a,b) and band structures (c,d)
for graphene over a fully in-plane triangular (a,c) and
rectangular (b,d) skyrmion lattice. A gap is opened
in the band structure, but the Chern number is identically zero,
yielding a trivial insulator.
}
\label{inplane}
\end{figure}

\section{Trivial gap opening}
In this section we will discuss some of the situations in which
a skyrmion texture will not give rise to a topological gap. 
\bluemark{First we emphasize that strong defects or inhomogeneities
both in graphene,
the underlying skyrmion lattice or the exchange coupling
can give rise to a phase transition between the topological
state to a trivial state. This problem is common to any
engineered topological state, and its study relies on the particular
features of the microscopic models, which depend on the underlying
material that hosts the skyrmions.}

In the following we focus on homogeneous effects well
captured by our phenomenological model that can ruin the perfect $\mathcal C=2$ state.
We will
discuss two situation, the case of a pure coplanar spin texture,
and situations with strong intervalley mixing.

\subsection{Coplanar spin textures}
Here we consider the case of graphene coupled to coplanar non-collinear spin textures, shown in Fig. \ref{inplane}.
We address the weak coupling limit ($J<<t$) at half filling.
In the case of a purely in-plane exchange field, the Hamiltonian
can be made real by a rotation onto the Pauli matrices $\sigma_x,\sigma_z$, leading to a
 vanishing  Hall response. We show in Fig. \ref{inplane}
examples of band structures of triangular and rectangular graphene
unit cells subjected to fully
in-plane exchange field. Although a band-gap opens,  
the calculated Berry curvature, and thereby the  Chern number vanish  in both cases. 

\bluemark{
The necessity of a non-coplanar spin texture can be easily understood in terms
of the symmetry properties of the Chern number. Without loss
of generality, the exchange field of
a coplanar spin texture
can be expressed in terms of the Pauli matrices $\sigma_x$ and $\sigma_z$,
provided the exchange field is rotated to lie in the $xz$ plane
\begin{equation}
\vec M \cdot \vec \sigma = M_x\sigma_x + M_z \sigma_z
\end{equation}
turning the exchange term of the Hamiltonian purely
real, and therefore
also the full Hamiltonian.
Since the Chern number is odd under conjugation
\begin{equation}
K : \mathcal C \rightarrow - \mathcal C
\end{equation}
and a purely real Hamiltonian is invariant under conjugation
\begin{equation}
K :  \mathcal{H} \rightarrow \mathcal{H^*} = \mathcal{H}
\end{equation}
the Chern number is identically zero ($\mathcal C=0$) for a co-planar spin texture.
Thus, non-coplanar spin arrangements are necessary to induce the anomalous Hall phase. 
}


  \begin{figure}[t!]
 \centering
 \includegraphics[width=0.5\textwidth]{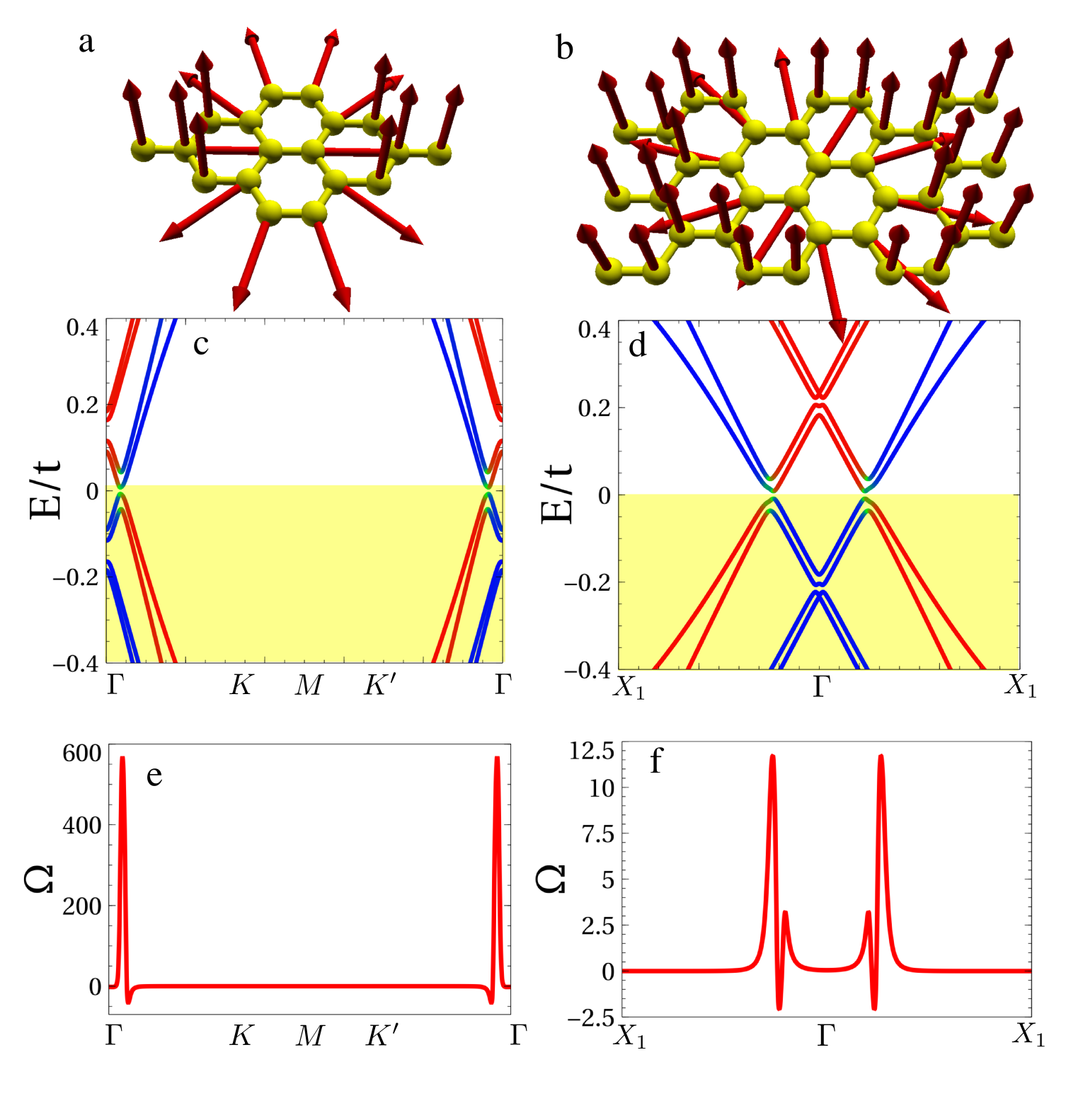}
\caption{Unit cells (a,b), band structures (c,d)
and Berry curvatures (e,f)
for graphene over a triangular (a,c) and
rectangular (b,d) skyrmion lattice.
The commesuration of graphene
with the skyrmion lattice make
that both valleys are folded onto the $\Gamma$ point. 
Even though a gap with non vanishing Berry curvature opens up, the sign
changes along the Brillouin zone, and summing up all the contributions
gives a vanishing Chern number. In the present situation,
the gap is dominated by intervalley mixing.
}
\label{folding}
\end{figure}

\subsection{Intervalley mixing}

In some instances the skyrmion lattice will be commensurate with the  carbon honeycomb lattice,
the two valleys in band structure of graphene
could be folded to a single point, allowing to intervalley
mixing. In particular, for the triangular lattice,  the
folding of the two valleys onto the $\Gamma$ point
takes place for $3n\times3n$ unit cells. For a rectangular unit cell,
the folding takes place for a supercell $3n$ in the zigzag direction.
In these situations, intervalley scattering can  open a trivial gap, as shown 
in Fig.\ref{folding} of such behavior. Interestingly,  the
Berry curvature can be non zero, but the  Chern number vanishes. 
Importantly,  this situation requires a fine-tuning of the skyrmion and carbon lattices.  In general, this is not the case, and 
non-trivial gaps are to be expected.


\end{document}